\documentclass[conference]{IEEEtran}
\usepackage{cite}
\usepackage{amsmath,amssymb,amsfonts}
\usepackage{algorithmic}
\usepackage{graphicx}
\usepackage{textcomp}
\usepackage{xcolor}
\usepackage{subfigure}
\usepackage{multirow}
\usepackage{mathtools}
\usepackage{enumitem}
\usepackage{booktabs}
\usepackage{color}
\usepackage{xcolor}

\def\BibTeX{{\rm B\kern-.05em{\sc i\kern-.025em b}\kern-.08em
    T\kern-.1667em\lower.7ex\hbox{E}\kern-.125emX}}
\begin{document}

\title{Retweets Amplify the Echo Chamber Effect}

\author{
Ashwin Rao, Fred Morstatter, and Kristina Lerman\\
Information Sciences Institute, University of Southern California\\
\{ashreyas, fredmors, lerman\}@isi.edu, 
}

\maketitle

\begin{abstract}
The growing prominence of social media in public discourse has led to a greater scrutiny of the quality of online information and the role it plays in amplifying political polarization. However, studies of polarization on social media platforms like Twitter have been hampered by the difficulty of collecting data about the social graph, specifically follow links that shape the echo chambers users join as well as  what they see in their timelines. As a proxy of the follower graph, researchers use retweets, although it is not clear how this choice affects analysis. Using a sample of the Twitter follower graph and the tweets posted by users within it, we reconstruct the retweet graph and quantify its impact on the measures of echo chambers and exposure.  While we find that echo chambers exist in both graphs, they are more pronounced in the retweet graph. We compare the information users see via their follower and retweet networks to show that retweeted accounts share systematically more polarized content. This bias cannot be explained by the activity or polarization within users' own follower graph neighborhoods but by the increased attention they pay to accounts that are ideologically aligned with their own views. Our results suggest that studies relying on the retweet graphs overestimate the echo chamber effects and exposure to polarized information.
\end{abstract}


\section{Introduction}
Social media's growing prominence as a platform for sharing news has raised concerns about the quality of information people see online and growing polarization.
Previous research 
focused on identifying accounts that promote misinformation~\cite{mabrey2021disinformation} or post extreme, emotionally charged content~\cite{carpenter2020political}. However, the influence of these accounts 
is complex and cannot be quantified simply by the volume of messages they generate, their number of followers or centrality within the follower graph. To understand the joint influence of online accounts on what people see, researchers have examined the role of networks in the formation echo chambers~\cite{conover2011political,cinelli2021echo,jiang2021social}, which link people to accounts that expose them to information that is congruent to their existing beliefs while isolating them from opposing viewpoints. By showing content that aligns with people's existing beliefs, echo chambers may increase polarization and reinforce harmful ideas.
 
Studies of online echo chambers vary in how they represent the network connections between users. Some 
(e.g., \cite{cinelli2021echo,nikolov2020right}) rely on the follower graph, 
because activity of \textit{``friends''}, i.e., accounts users follow, largely determines what people see. 
However, constructing echo chambers from the follower graph is highly non-trivial due limitations of the Twitter API and the number of queries required to collect all links. As a result, the friend/follower links remain largely unobserved. 
On the other hand, retweeting,  which refers to the practice of re-sharing another user's posts, can be easily observed in the tweet metadata returned by the API or in the text of the tweet. 
Several studies have, therefore, used retweet links to study echo chambers~\cite{conover2011political,jiang2021social,sasahara2021social}.
To date, however, there is little understanding of how well the retweet links capture the information people 
receive from accounts they follow, 
their relationship to echo chambers in the follower graph, and the biases they introduce in the analysis of the information exposure. 

In this work we compare network neighborhoods in the follower and retweet graphs
and explore how they affect our estimates of what people see on Twitter. 
We organize our research around the following questions:

\begin{description}
\item[RQ1] What is the relationship between follower-graph friends (i.e., accounts a user follows) and retweet-graph friends (i.e., the accounts the user retweets)? 
\item[RQ2] Do echo chambers exist in both the retweet and follower graphs? If so, what are the differences? 
\item[RQ3] Are there systematic differences in information exposure via follower-graph and retweet-graph friends?
\item[RQ4] What accounts for these differences?
\end{description}

To answer these questions we leverage a dataset containing comprehensive information about more than 5K Twitter users, i.e., \textit{seed users}, and all their follower-graph friends. The links between seed users and their friends create a subset of the Twitter follower graph. The dataset also contains all  messages that seed users and their friends posted over a period of six months. At the time of data collection, Twitter had no algorithmic curation and displayed tweets in reverse chronological order to users. This allowed us to reconstruct the information seed users saw in their timelines, i.e., their \textit{information exposure}. We also extracted links to the accounts seed users retweeted, i.e., their \textit{retweet-graph friends}.  
Aggregating messages shared by retweet friends creates a proxy for information exposure (in the absence of algorithmic curation). 

We quantify a user's information exposure by assessing the political ideology of content the user sees, specifically how hardline it is (either hard right and hard left). 
We show that \textit{echo chambers exist in both types of graphs}: 
users are connected to 
accounts ideologically congruent views, who expose them to similar information to what they themselves share. 
However, retweet friends are systematically skewed: ideologically extreme users are connected to more extreme retweet friends. As a result, our estimates of information exposure are also systematically distorted: ideologically extreme users see more polarized information via their retweet-graph friends compared to their follower-graph friends. 
This bias can be explained by users preferentially paying more attention to  more extreme content.

Our results suggest that studies of online echo chambers based on the retweet graphs may overestimate the polarization of information users see online. Moreover, while even ideologically extreme users follow a variety of accounts, they appear to selectively pay attention to the more extreme of these accounts. 
Mitigating polarization and exposure to extreme  information will require measures beyond reducing online echo chambers.

\section{Related Works}

Previous studies have leveraged follow relationships to infer individual political preferences \cite{barbera2015tweeting}, assess characteristics of echo chambers \cite{garimella2017long,bakshy2015exposure,barbera2015tweeting} and quantify ideological exposure \cite{barbera2015tweeting,himelboim2013birds,dyda2019hpv}. The common assumption  of these studies is that users are more likely to follow others who are ideologically similar to them. Studies of echo chambers on social media platforms like Twitter relied on retweet and mention interactions between individuals \cite{stewart2018examining,conover2011political}. While some studies found strong ideological clustering \cite{cossard2020falling,garimella2017long,himelboim2013birds, conover2012partisan}, others have highlighted the existence of cross-ideological exposures \cite{bakshy2015exposure,barbera2015tweeting}. A recent study  compares the misinformation exposure in co-follower and co-retweet networks for followers of political elites~\cite{mosleh2022measuring}.

In order to better understand echo chamber effects one needs to better characterize exposures. The proliferation of content generated on social media has brought with it an overload of information. A survey based experiment \cite{bontcheva2013social} showed the users of micro-blogging platforms like Twitter are the worst affected with nearly two-thirds of the users feeling overloaded with information. Studies on Twitter and Sina Weibo \cite{feng2015competing,rodriguez2014quantifying} have found that users who have many friends needed repeated exposures to the same content before they re-shared it. These highlight the importance of factoring in the user attention span while characterizing exposures. One way to do so is by directly looking at the content re-shared (or retweeted on Twitter) by individuals as an unified abstraction of such repeated exposures. Moreover, a follow relationship between two users need not necessitate ideological similarity and could arise out of mere curiosity. While several studies have explored factors affecting retweetability of content and retweet information cascades \cite{suh2010want,pulido2020covid,petrovic2011rt,zaman2010predicting}, not many of them have explored their role in user exposures. While comparisons of the structure of follower and retweet networks have been done before \cite{bild2015aggregate}, a comparison of exposures from neighborhoods in these networks has remained unexplored. 

The growing influence of content curation algorithms on user timelines has motivated research in understanding how content exposures are affected \cite{huszar2022algorithmic,bartley2021auditing}. There is an active debate on how one can address echo chambers \cite{spohr2017fake}. One viewpoint argues that cross-ideological exposure can mitigate echo chambers \cite{dubois2018echo} while others have argued that the control over exposure to cross-ideological content lies with the user themselves \cite{bakshy2015exposure}. This finding suggests the presence of selective attention and motivates us to quantify exposures in the absence of influence from recommendation systems to better understand the dynamics of what users pay attention to. In other words, do users  selectively attend to some information in their timelines and if so, how can retweet exposures be leveraged to understand this.


\begin{figure}
    \centering
    \includegraphics[width=0.98\linewidth]{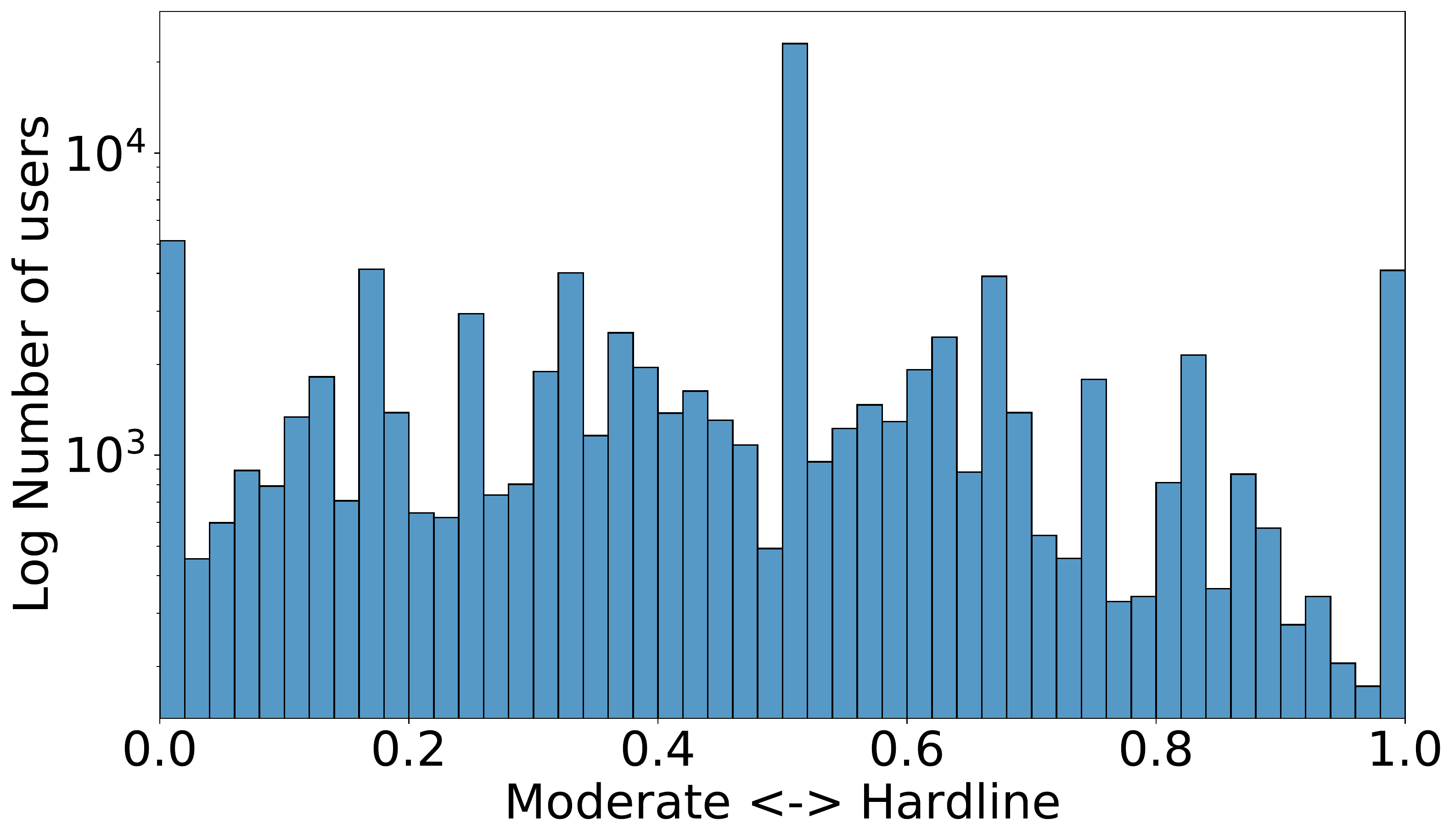}
    \caption{Frequency distribution of political moderacy scores of seed users. Scores near zero are associated with politically centrist or moderate content while scores near one are associated with politically hardline content, either hard right or hard left.}
    \label{fig:user_scores}
\end{figure}

\section{Data and Methods}

\begin{figure*}[tb]
    \centering
    \subfigure[Fraction of follower-graph friends retweeted]{\includegraphics[width=0.32\linewidth]{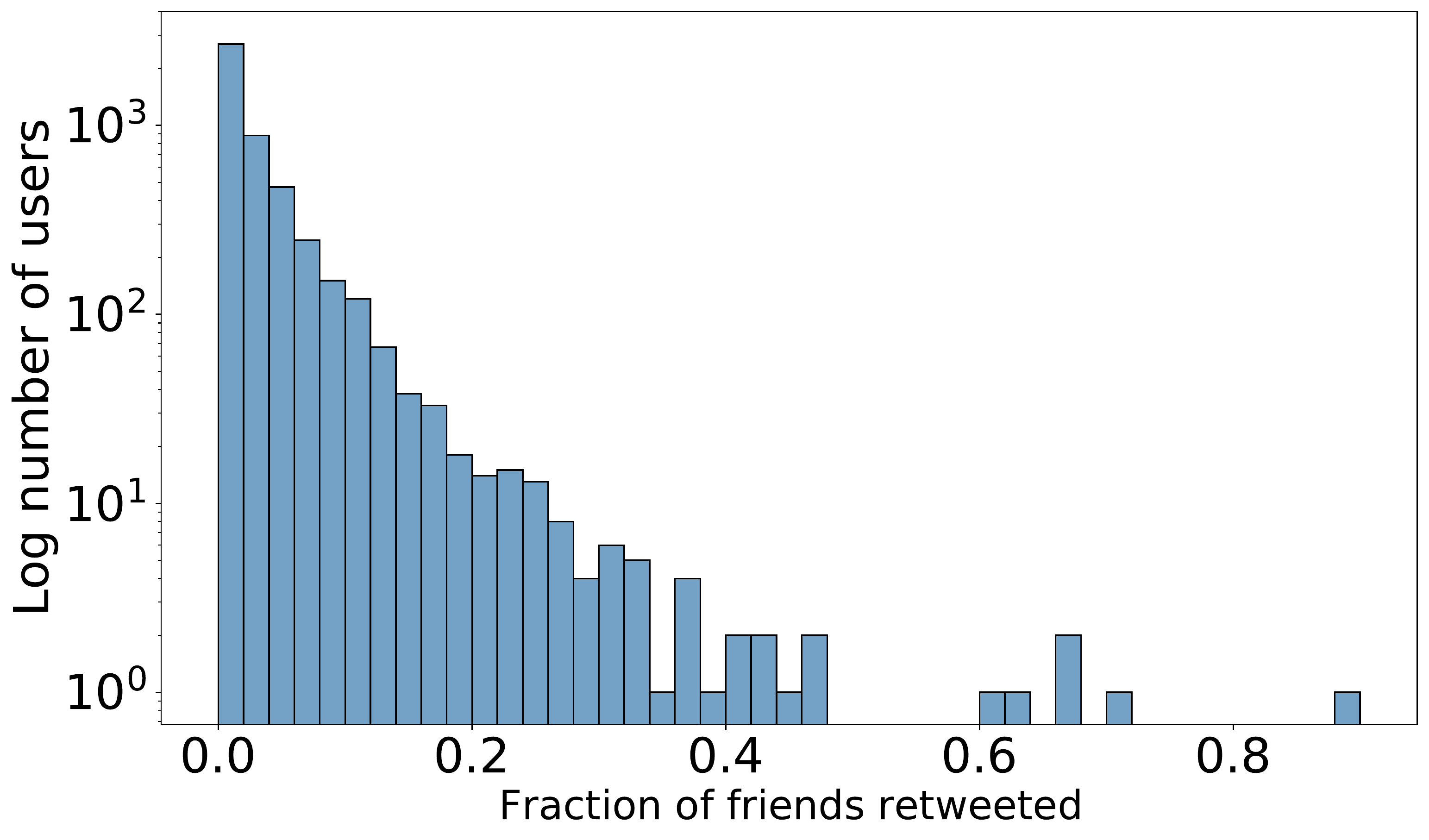}}
    \subfigure[Overlap of retweet-graph friends and follower-graph friends]{\includegraphics[width=0.32\linewidth]{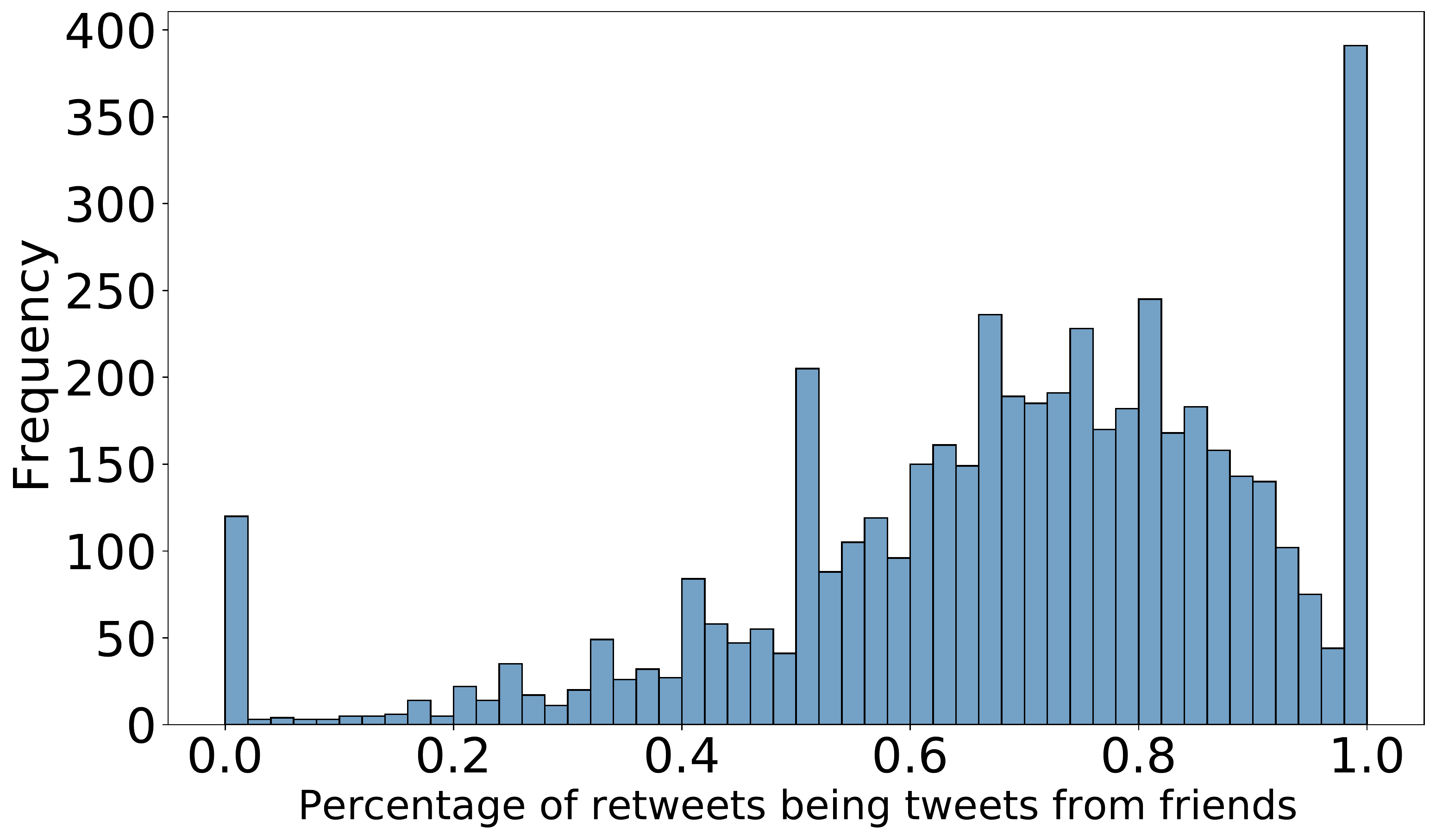}}
    \subfigure[Overlap change as retweet interactions increase]{\includegraphics[width=0.32\linewidth]{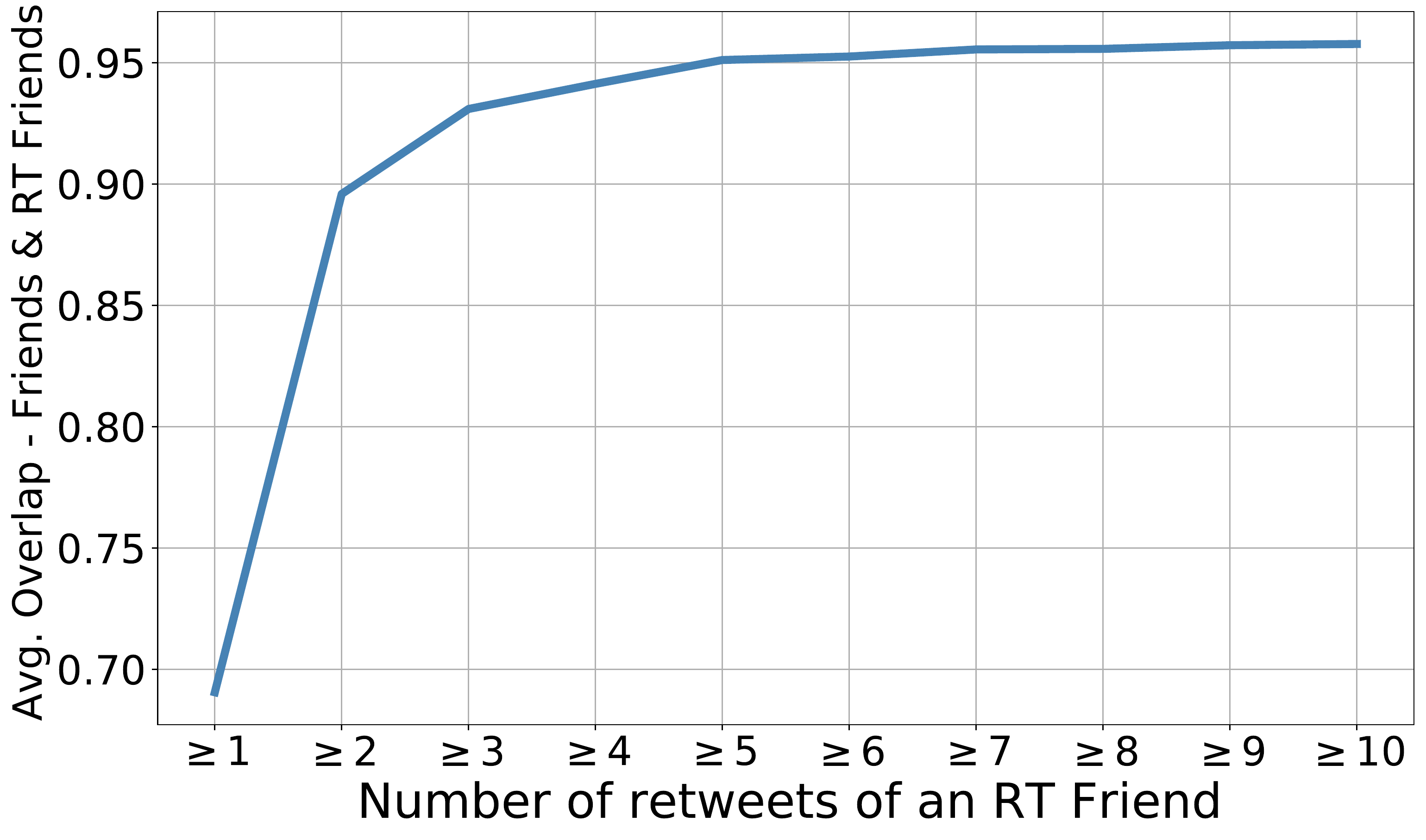}}
    \caption{Relationship between local structure within the follower graph and the retweet graph. (a) Distribution of fraction of follower-graph friends who were retweeted. (b) Distribution overlap between who individuals follow and retweet. (c) As the minimum number of retweet interactions increase between a user and a retweet friend, the overlap between follower-graph friends and retweet-graph friends increase.}
    \label{fig:overlap}
\end{figure*}

Data collection was based on a study that tracked discussions of initiatives on the 2012 California ballot~\cite{smith2013role}. These initiatives proposed new laws on topics such as labeling genetically modified foods, abolishing the death penalty, and school funding. The study identified 81 users active in the discussion of ballot initiatives and used snowball sampling to expand this set to $5,599$ accounts, who we call \textit{seed users}. 

\textbf{Follower graph:}
Starting in March 2014, we queried Twitter for the accounts each seed user follows, who we call \textit{friends} or \textit{follower graph friends}. We queried daily through September 2014 to identify any new friends. This subset of the Twitter follower graph has over 4M users with over 17M edges.

We also collected messages seed users and their friends posted over this time period, roughly 81.2M tweets with 22.7M retweets. At the time of data collection, Twitter showed tweets posted by friends in reverse chronological order in a user's timeline. We were therefore able to reconstruct the timeline for each seed user and quantify information exposure. For this study we consider tweets from May--September 2014, or 43.4M tweets of which 14.8M are retweets.

\textbf{Retweet graph:}
We identified retweets (RTs) posted by seed users and created links from  accounts they retweeted. The \textit{retweet graph}  aggregates retweet links over all seed users. The tweet object specifies whether it is a retweet and gives a link to the account who originally posted it. Intermediate retweet chains are not recorded in the tweet object. We did not collect all tweets generated by retweet friends in cases where the retweet friend was \textit{not} a friend of a seed user.

\textbf{Limitations:}
Note that the data set has some limitations. The data was collected before Twitter algorithmically personalized timelines in 2016, so it does not reflect how users are exposed to information now. However, this enables us to study the impact of networks on the information users see without the confounding effects of algorithms. Also, Twitter still allows users to select to see tweets in reverse chronological orders and other social platforms, such as Mastodon and Instagram, also allow content to be shown in reverse chronological order. This adds to our study's relevance.

Another limitation that seed users set has a liberal bias. Most of the people discussing ballot initiatives in California election are liberal, which contributes to the dearth of conservatives in our sample. Despite these limitations, we believe that this unique data offers an unprecedented opportunity to study exposure in online social networks. Instead of focusing on understanding exposures in the conventional eyes of ideology, we focus on moderacy which captures the intensity of ideological belief on a scale from moderate($0$) to hardline ($1$).  

\subsection{Measuring Polarization}

Following previous studies \cite{le2019measuring,nikolov2020right,cinelli2021echo,rao2021political}, we quantify ideology based on URLs users share in their original tweets.  Media Bias-Fact Check (http://mediabiasfactcheck.com) (MBFC) scored thousands of Pay-Level Domains (PLDs) along multiple dimensions, including partisan bias, political moderacy and quality of information they share.   
MBFC categoritzes ideological slant of a PLD as Left/Hardline Liberal (which we score $0$), Left-Center ($0.25$), Least-Biased/Center ($0.5$), Right-Center ($0.75$), Right/Hardline Conservative ($1$). 
In our sample of users, roughly half, or 2.8K seed users, generated ideological content with URLs to these PLDs. This content was skewed towards liberals; therefore, for a more balanced distribution, we focus on the \textit{moderacy} dimension in our analysis. 

\subsubsection{Individual Moderacy} 
Moderacy is a continuous score 
representing the intensity of political ideology, ranging from centrist/least biased ($0$) to hardline  ($1$),  either Hardline Liberal or Hardline Conservative.
To get a user $u$'s moderacy score $m_s(u)$, we calculate the weighted average of domain scores $\Pi(d)$ of the URLs embedded in $u$'s tweets. If the computed weighted average $m_s(u) \leq 0.5$, we subtract the value from 1. Given a set of domains $D(u)$  user $u$ shares, $m_s(u)$ is:

$$
m_s(u)=\begin{cases}
        \frac{1}{|D(u)|}\sum_{d \in D(u)} \Pi(d), & \text{if } m_s(u) > 0.5\\
        1 - \frac{1}{|D(u)|}\sum_{d \in D(u)} \Pi(d), & \text{if } m_s(u) \leq 0.5\\
    \end{cases}
$$
\noindent
We rescale the scores to $[0,1]$ range using min-max normalization. 
The distribution of individual moderacy scores, Fig.\ref{fig:user_scores}, shows that our user sample is nearly uniformly distributed along the moderacy dimension.  

\subsubsection{Moderacy of Exposures}
Previous works \cite{cinelli2021echo,nikolov2020right} quantified polarization of a user's neighborhood, i.e., the echo chamber effect, by averaging over each friend's political (or factual) orientation. However, this ignores the large variation of friends' activity, with each friend contributing equally to neighborhood polarization, regardless of how many messages the friend posts. In contrast, we estimate 
information exposure by aggregating all tweets the user's follower-graph friends (resp. retweet-graph friends) shared. This gives more weight to the more active friends. Let us denote the relationship between two nodes using $\rho$ where, $\rho \in \{friend, retweet~~friend\}$. $D(\rho)$ denotes the domains shared by a user's follower-graph friends (or retweet-graph friends).
The moderacy of \textit{exposure} is:
$$
    m_e(u,\rho) = \begin{cases}
        \frac{1}{|D(\rho_u)|}\sum_{d \in D(\rho_u)} \Pi(d), & \text{if } m_s(u) > 0.5\\
        1 - \frac{1}{|D(\rho_u)|}\sum_{d \in D(\rho_u)} \Pi(d), & \text{if } m_s(u) \leq 0.5\\
    \end{cases}
$$

We denote exposures via the follower graph $(\rho = f)$ as $m_e(u,f)$ and for the retweet graph $(\rho=r)$ as $m_e(u,r)$. We also rescale the scores to $[0,1]$ range. 

\section{Results}

\begin{figure}[th]
    \subfigure[Follower Graph Exposure]{\includegraphics[width=0.98\linewidth]{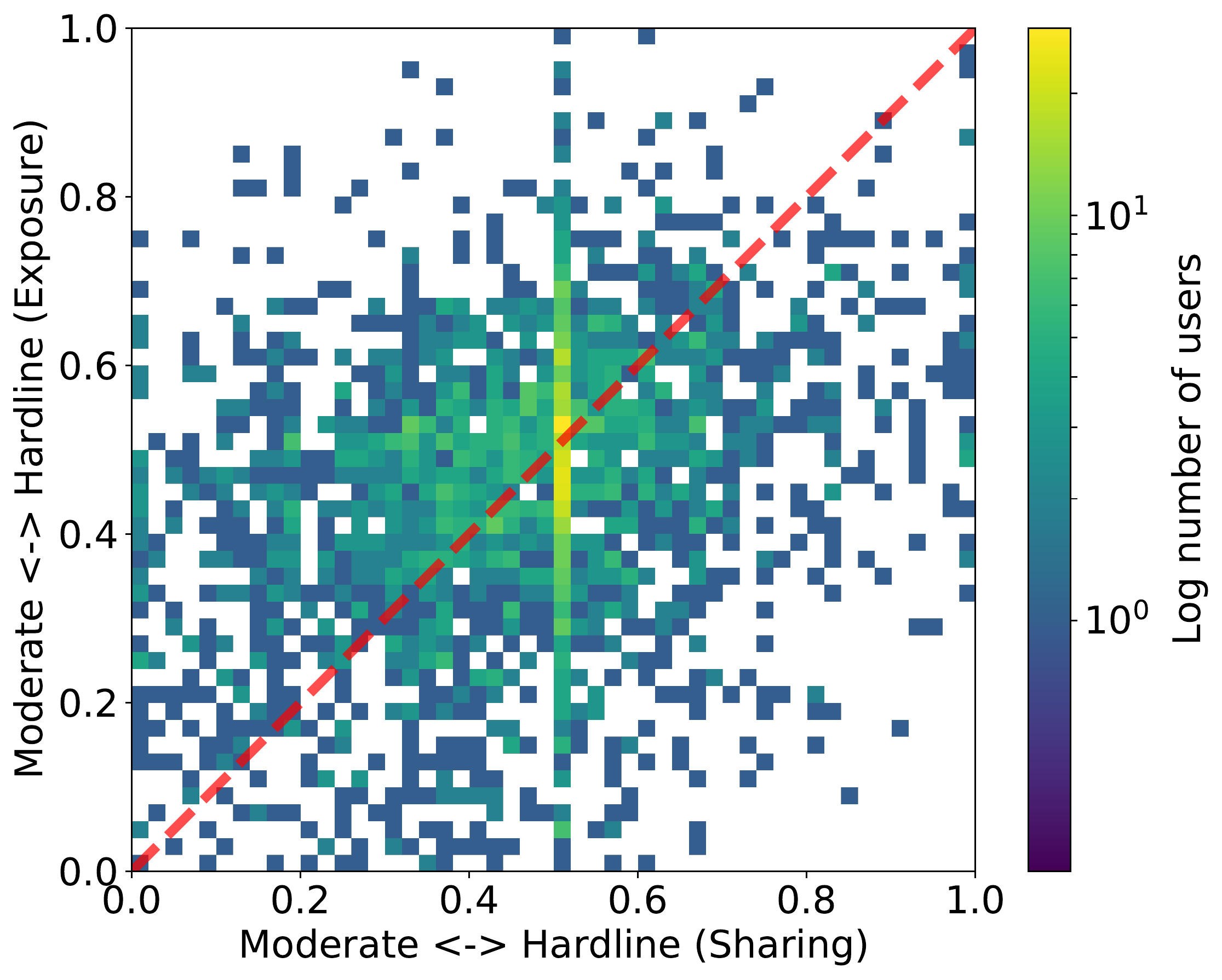}}
    \subfigure[Retweet Graph Exposures]{\includegraphics[width=0.98\linewidth]{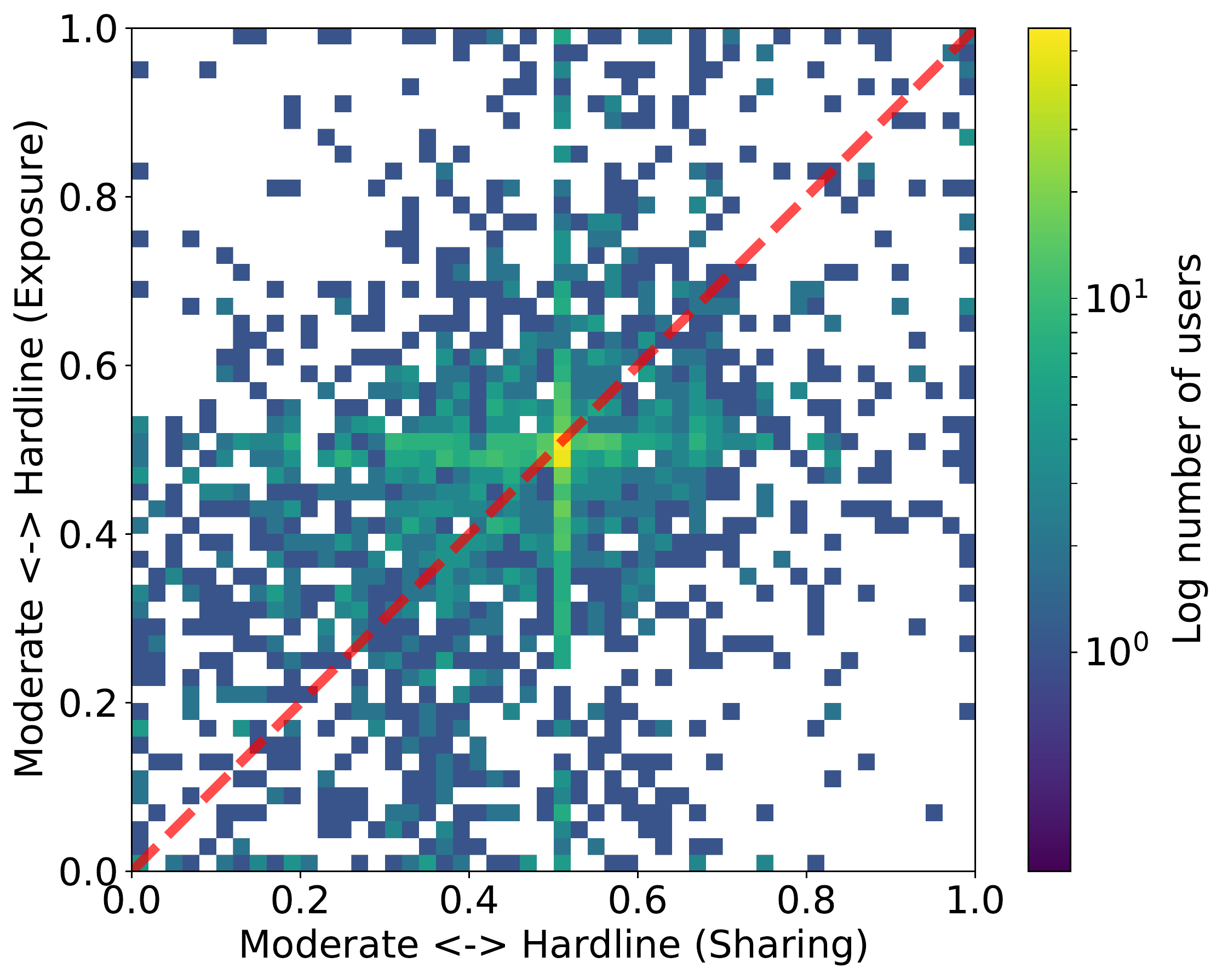}}
 \caption{Heatmap of the number of seed users with given individual moderacy $m_s$ and moderacy of exposure $m_e$ via the (a) follower-graph  and (b) retweet-graph friends. Correlations  are $0.26 (p<0.001)$ and $m_s$ and $0.31 (p<0.001)$ in (a) and (b) respectively.}
\label{fig:echo_plots}
\end{figure}

\begin{figure*}[th]
    \subfigure[Follower Network]{\includegraphics[width=0.33\linewidth]{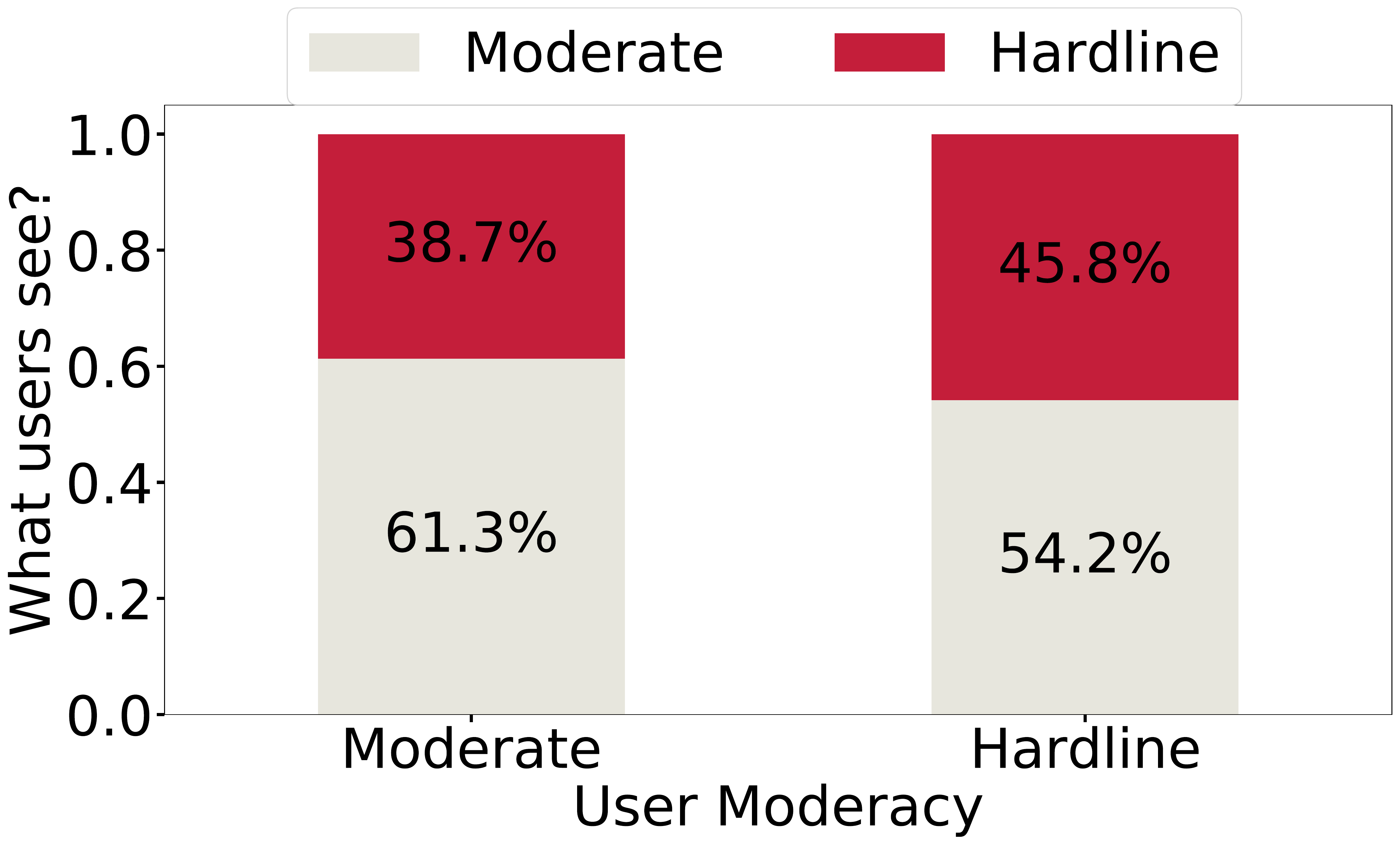}}
    \subfigure[Retweet Network]{\includegraphics[width=0.33\linewidth]{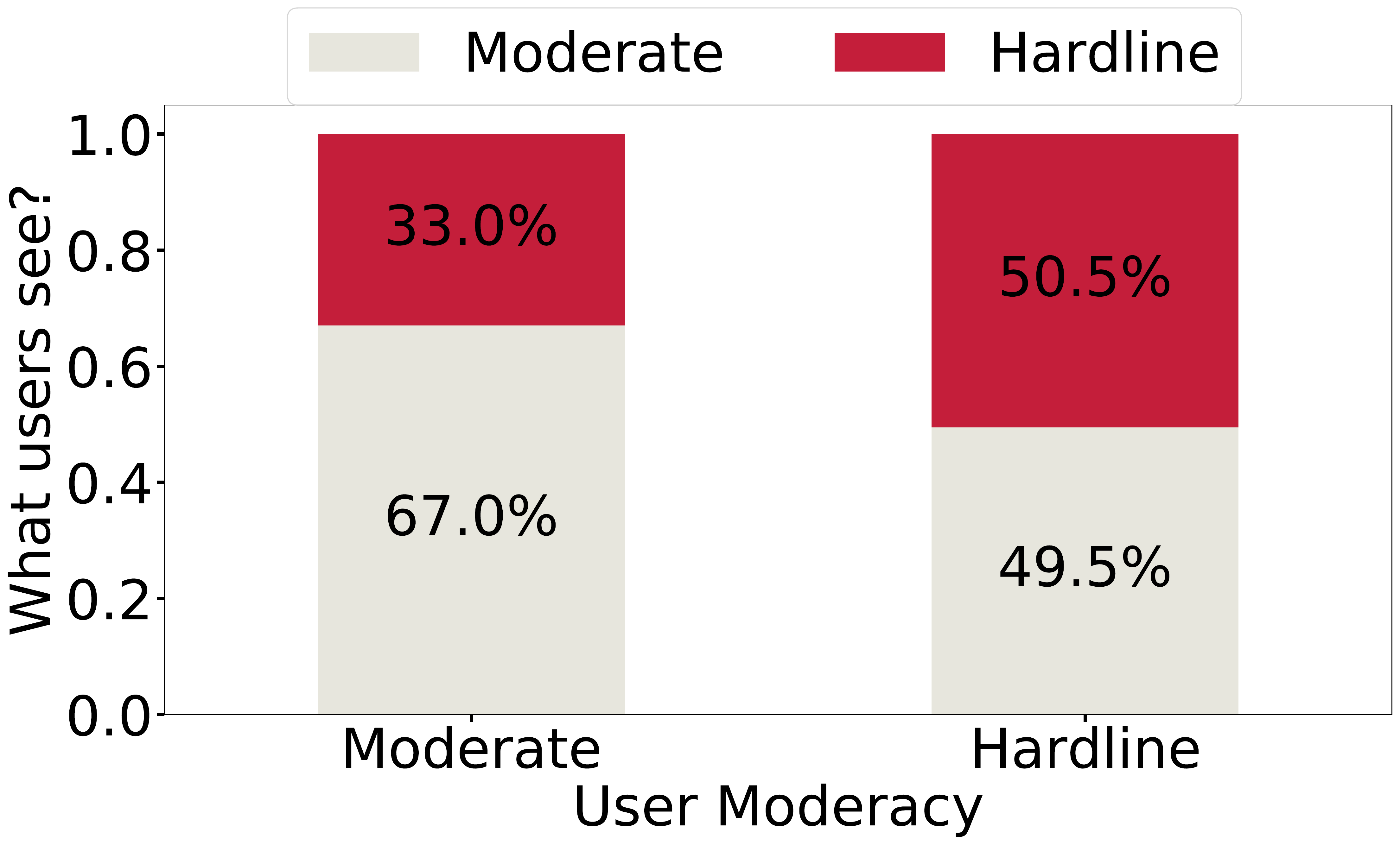}}
    \subfigure[Random Network]{\includegraphics[width=0.33\linewidth]{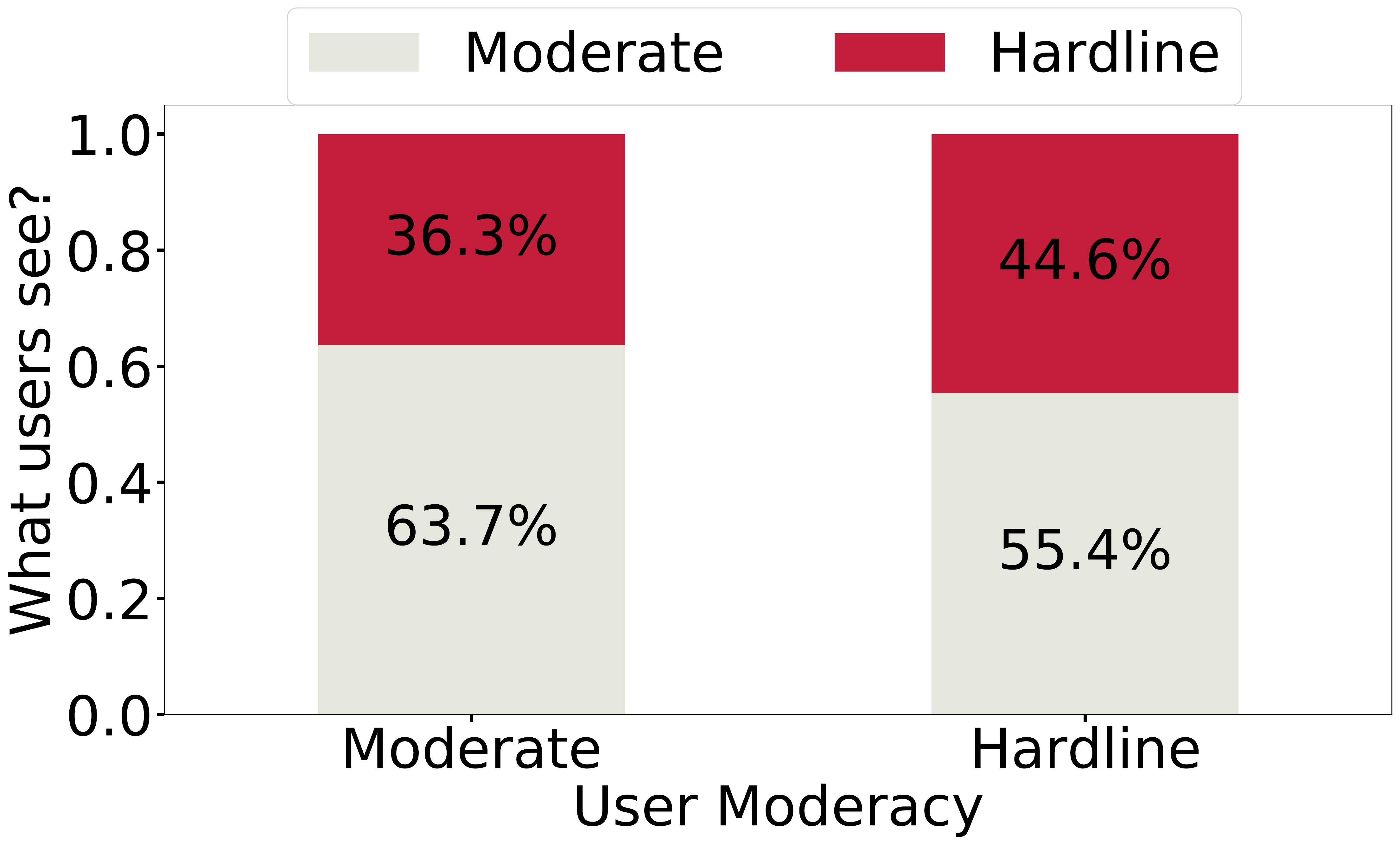}}
 \caption{Exposure to polarized information from follower-graph friends (a), retweet-graph friends (b) and a random subset of follower graph friends (c). Retweet-graph friends expose users to content that is most aligned with the users' polarities.}
 
\label{fig:exposure_vs_sharing}
\end{figure*}

\subsection{Follower-graph Friends vs Retweet-graph Friends}

First, we compare user's number of friends in the follower graph and the retweetgraph. As a reminder, we define a \textit{retweet-graph friend} (or \textit{retweet-graph friend}) of a seed user as the account the seed user has retweeted $k$ or more times, where $k$ is a threshold we vary from $1$ to $10$. 

We calculate the share of follower-graph friends who have been retweeted at least once by each seed user. Fig. \ref{fig:overlap}(a) shows the distribution of this quantity. 
Retweet-graph friends are a sparse approximation of the follower graph: roughly $90\%$ of the seed users retweet fewer than $10\%$ of their follower-graph friends. 

Next, we assess the overlap between follower-graph friends and retweet-graph friends: what fraction of a seed user's retweets are of their follower-graph friends' content.  On average the overlap between users' follower-graph friends and retweet-graph friends is $0.69$, and for roughly half of all seed users, $72\%$ of the content they retweet comes from their follower-graph friends (Fig. \ref{fig:overlap}(b)). 

To study how  users distribute their attention over follower-graph friends, we look at how the overlap between follower-graph friends and retweet-graph friends changes as we increase the threshold that defines the minimum number of times a seed user needs to retweet an account for it to be considered a retweet-graph friend. As we increase the threshold from $1$ to $10$, we find that users retweet a higher fraction of their follower-graph friends' tweets. Figure \ref{fig:overlap}(c) shows the change in average overlap between follower-graph friends and retweet-graph friends as a function of threshold. 
When the threshold is $2$, meaning a seed user has to retweet an account at least twice to be counted a retweet friend, the average  overlap between follower-graph friends and retweet-graph friends  is $0.89$. These results suggest that users give their follower-graph friends sustained attention. Unless otherwise stated, we consider at least one retweet interactions for an individual to be considered a user's retweet friend.  

These findings answer our first question: \textit{The retweet-graph is a sparse approximation of the follower graph, with users paying sustained attention to a small subset of their friends.}

\subsection{Follower-Graph Exposure vs Retweet-Graph Exposure}
Before Twitter introduced algorithmic curation, users selected which accounts to follow and saw the content from those accounts in their timelines. In our pre-curation data, follower-graph  friends, therefore, shaped the information users saw online. This enables us quantify how using the retweet graph as a proxy of follower graph changes our observations of echo chambers and estimates of information exposure. 


\subsubsection{Echo Chambers} 
By following accounts with similar views, users create echo chambers that expose them to information compatible with pre-existing opinion, while isolating them from opposing viewpoints. However, some of the previous studies relied on the follower graph to measure the phenomenon~\cite{cinelli2021echo,nikolov2020right}, while others leveraged retweet networks~\cite{conover2011political,jiang2021social,sasahara2021social}.
Here, we compare these different representations of echo chambers focusing on the moderacy dimension of political ideology. 
Figure~\ref{fig:echo_plots} shows the relationship between the individual user moderacy $m_s(u)$ and the moderacy of exposure via the follower-graph friends (a) and retweet-graph friends (b) as a density plot. There is a weak (but statistically significant) correlation between these measures, indicating that seed users who share content with specific moderacy scores tend to be exposed to similarly-valent content, i.e., echo chambers exist. There is a stronger correlation between individual moderacy $m_s$ and exposure via retweet friends (Pearson's $r=0.31, p<0.001$) compared to exposure via the follower-graph friends (Pearson's $r=0.26, p<0.001$). 

To further demonstrate the echo chamber effect within the retweet and follower graphs, we binned seed users into two groups based on their moderacy scores: Moderates ($m_s \leq 0.5$), and Hardliners ($m_s > 0.5$). Our analysis revealed a total of $63.2K$ moderates, and $29.3K$ hardliners. We then examined the proportion of moderate and hardline tweets that these seed users were exposed to via their follower-graph and retweet-graph connections. The results, depicted in Figs. ~\ref{fig:exposure_vs_sharing} (a) and (b) clearly demonstrate that retweet-graph connections expose moderate and hardline users to a higher percentage of ideologically congruent content compared to follower-graph connections. Amplification of content with similar moderacy attitudes is particularly evident in retweet-graph exposures (Fig. ~\ref{fig:exposure_vs_sharing}(b)) compared to follower-graph exposures (Fig. ~\ref{fig:exposure_vs_sharing}(a)), indicating a stronger echo chamber effect within the retweet network. Specifically, for hardline users, about $51\%$ of their exposures via their retweet-graph friends is also hardline as compared to about $46\%$ from their follower-graph friends. Similarly, $67\%$ of the exposures from retweet-graph friends for moderate users is also moderate as opposed to $61\%$ from follower-graph friends.

In order to assess the reliability of this comparison, we randomly select friends from a users' follower-graph connections equal to the number of retweet friends they have and calculate the fraction of moderate and hardline content these random set of friends expose the user to. We repeat this process 1000 times. Fig. \ref{fig:exposure_vs_sharing}(c) shows the average, over 1000 runs, moderate and hardline content that these random sets of friends expose users to. We find that retweet-graph friends expose users to the most ideologically identical content and as such cannot be attributed to random noise.

\begin{figure}
    \centering
   \subfigure[Distributions of Moderacy Scores]{\includegraphics[width=0.98\linewidth]{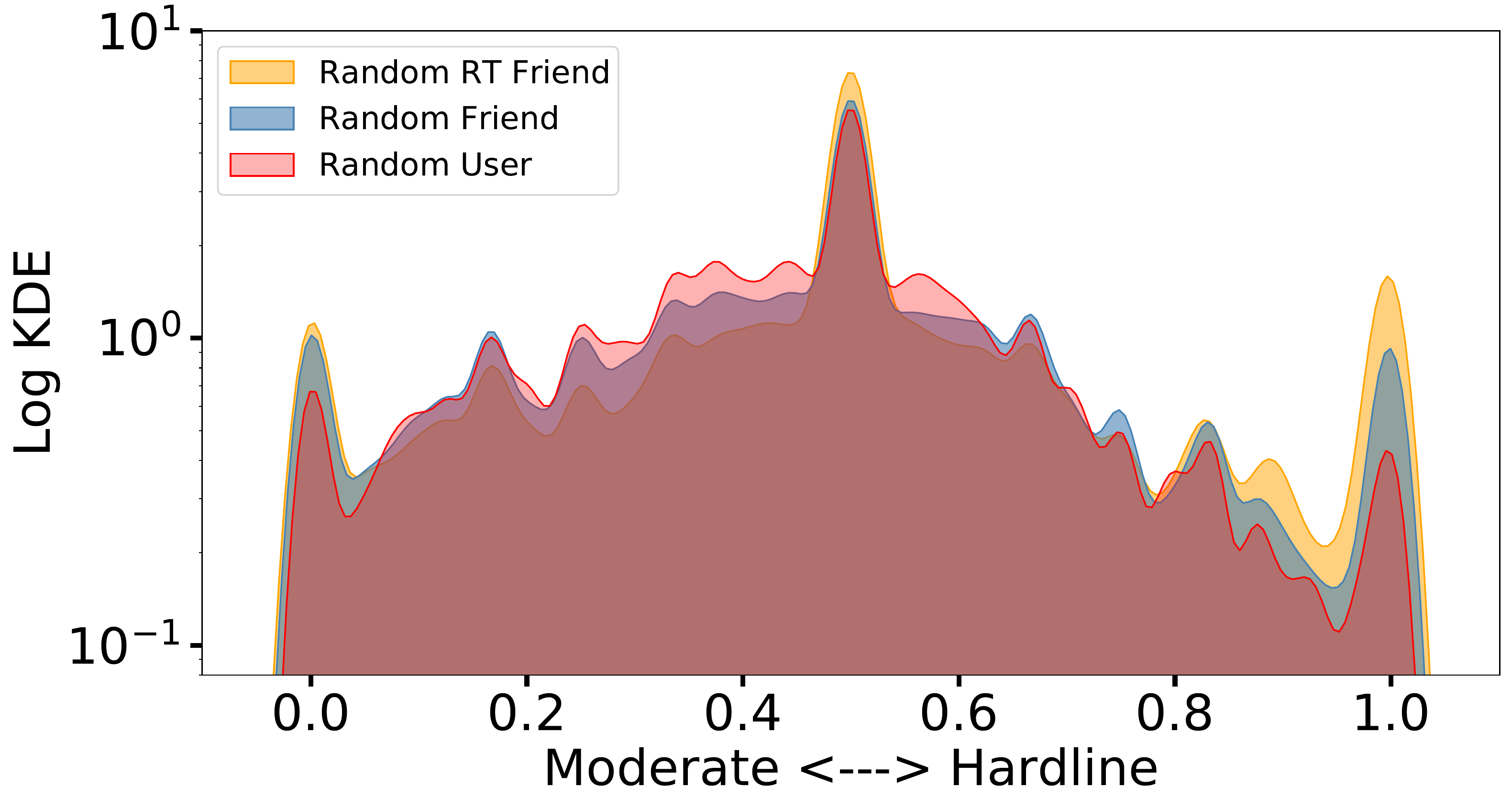}}   
   \subfigure[Shannon Entropy]{\includegraphics[width=0.98\linewidth]{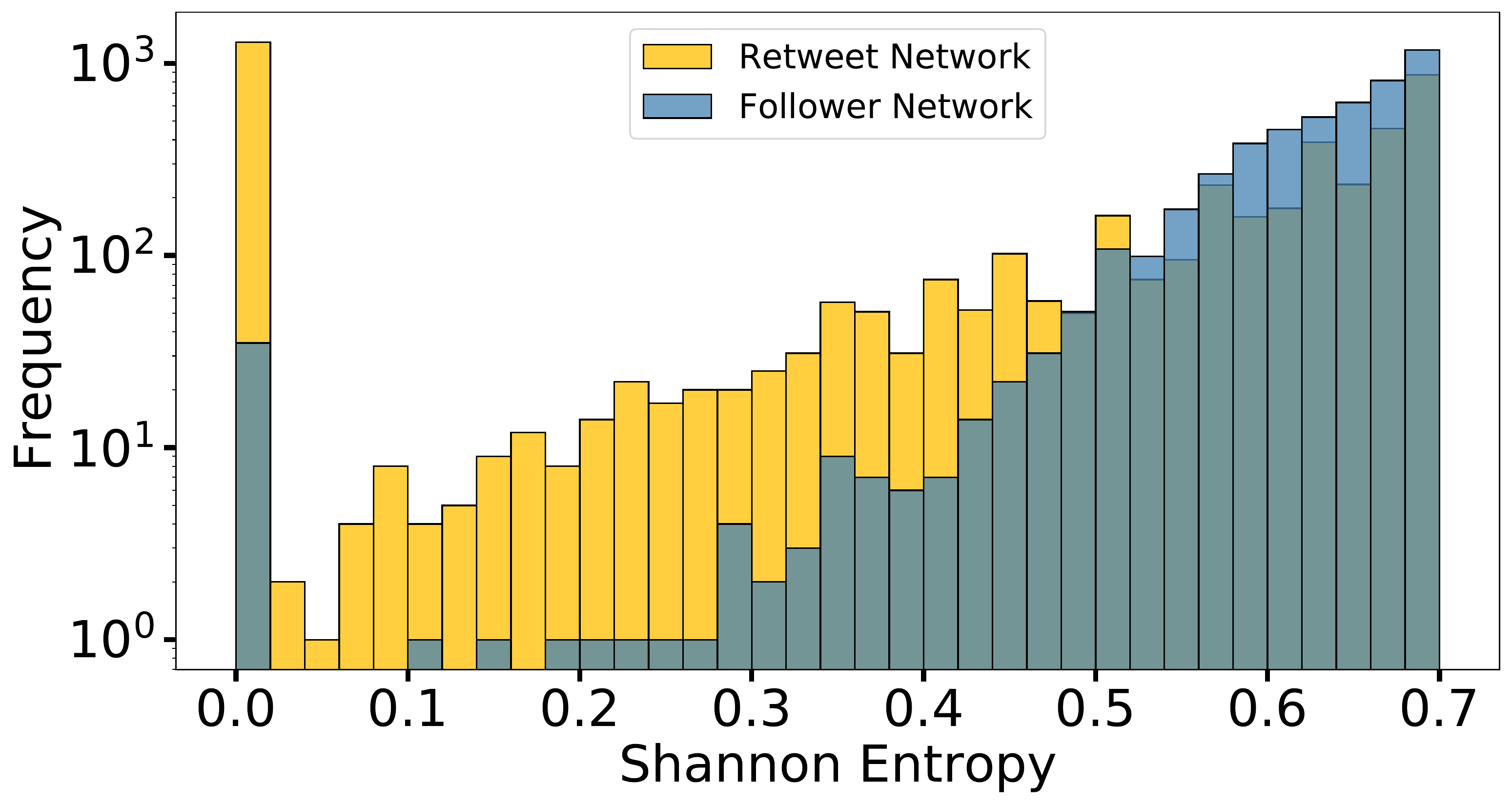}}
    \caption{(a) Compares the distributions of moderacy scores of random user, random follower-graph friend and a random retweet-graph friend($m_s$). The y-axis shows the kernel density estimate on a log-scale. (b) Shannon Entropy of polarities of users' retweet-graph and follower-graph friends. Retweet-graph friends of users have less diverse polarities in comparison to users' follower-graph friends. Differences are statistically significant under Mann-Whitney U Test at $p<0.001$. }
    \label{fig:friends_entropy}
\end{figure}

We randomly sample (N=500,000) follower-graph and retweet-graph friends of users proportional to their indegree in the follower and retweet networks respectively. More specifically, friends and retweet-friends having a high indegree have a higher propensity to be selected in the sampling process. We refer to them as random friends and random retweet friends. We sample individual user moderacy scores using as a uniform distribution to get the distribution of random user score. Figure ~\ref{fig:friends_entropy}(a) compares the distributions of moderacy scores of random users, random friends and random retweet friends. We find that scores for random retweet friends are systematically higher at extremes than scores of friends.

Additionally, we calculate the Shannon entropy of ideologies of users' retweet-graph and follower-graph friends. This assesses whether users pay attention to a diverse set of friends or prefer ideologically homogeneous friends. The lower the Shannon entropy, the less ideologically diverse the set of friends. Fig. \ref{fig:friends_entropy}(b) compares the distribution of entropy values of retweet-graph and follower-graph friends. We see that retweet-graph friends are less ideologically diverse than follower-graph friends. The difference is statistically significant under the Mann-Whitney U Test ($p<0.001$). 


\begin{figure*}[th]
    \centering
    {\includegraphics[width=0.98\linewidth]{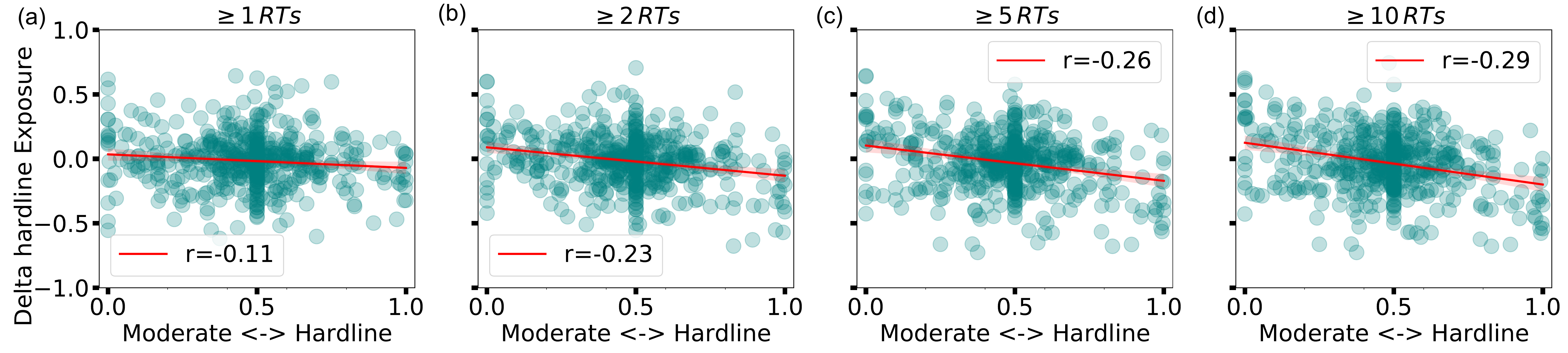}}
    \caption{Retweets amplify exposure to ideologically congruent information. As the overlap between retweet friends and friends increase with an increase in frequency of retweet interactions between a seed user and their retweet friend, we see an increased distortion in exposures. (a--d) Difference in moderacy exposure $\Delta m_e = m_e(u,f)-m_e(u,r)$ as a function of seed user's moderacy score $m_s(u)$ as we increase the threshold defining how many times a seed user retweets an account for that account to be counted as a retweet friend.}
    \label{fig:echo_threshold}
\end{figure*}

\subsubsection{Exposures via Retweet Graph are Systematically Biased}

Next, we demonstrate that content shared by retweet-graph friends is systematically 
biased.
To quantify, we define $\Delta m_e(u) = m_e(u,f) - m_e(u,r)$, the difference in the moderacy of tweets shared by follower-graph friends and retweet-graph friends. Figs. \ref{fig:echo_threshold}(a)--(d) show $\Delta m_e$ as a function of seed user's moderacy score $m_s$. The correlation between $\Delta m_e(u)$ and $m_s$ is negative, which means that 
retweet friends share systematically more hardline information than follower-graph friends, and this discrepancy is larger for more hardline seed users. In other words,
as individuals become more hardline (resp. moderate), their retweet-graph friends share even more hardline (resp. moderate) content than their follower-graph friends post. Therefore, using follower graphs to calculate exposure underestimates the extent to which individuals pay attention to ideologically congruent information. The magnitude of bias increases with the weight of retweet links: 
as we increase how much attention users pay to friends (by varying the retweet threshold), exposure bias increases. As we raise the retweet threshold in Fig.~\ref{fig:echo_threshold}(a)-(d), the Pearson's correlation decreases from $r=-0.10 (p<0.001)$ threshold of $\geq 1$ to $r=-0.29 (p<0.001)$ for threshold of $\geq 10$. 

These results 
answer our second and third research questions: echo chambers in the retweet graph are somewhat stronger than in the follower graph. Users, especially more hardline users, selectively retweet others with similar views. 
As a result, retweet graph may exaggerate the echo chamber effect that exists in the follower graph.

\begin{figure}[th]
    \subfigure[Activity: Retweet Friend vs Friend]{\includegraphics[width=0.49\linewidth]{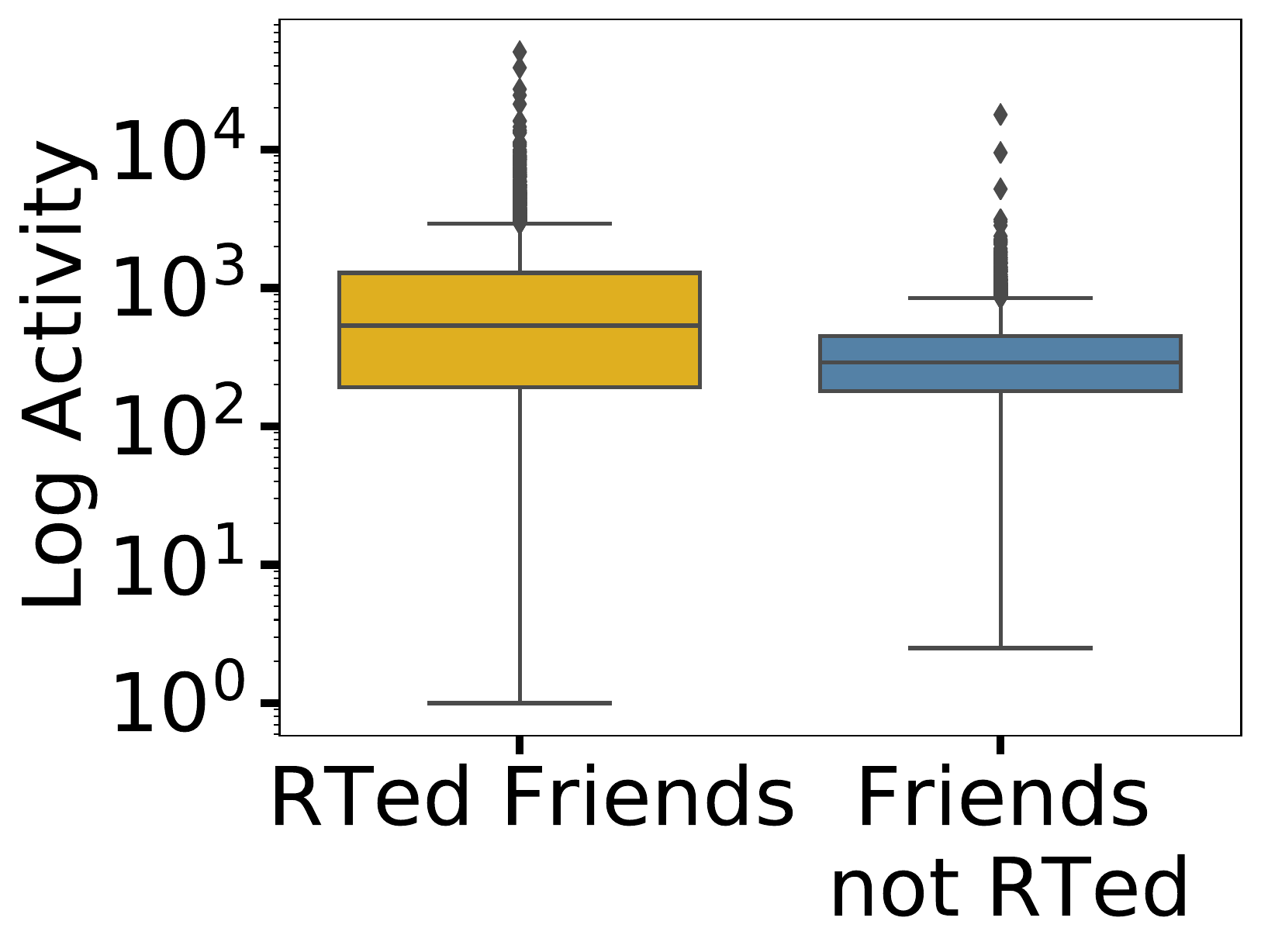}}
    \subfigure[Activity of retweeted follower-graph friends]{\includegraphics[width=0.49\linewidth]{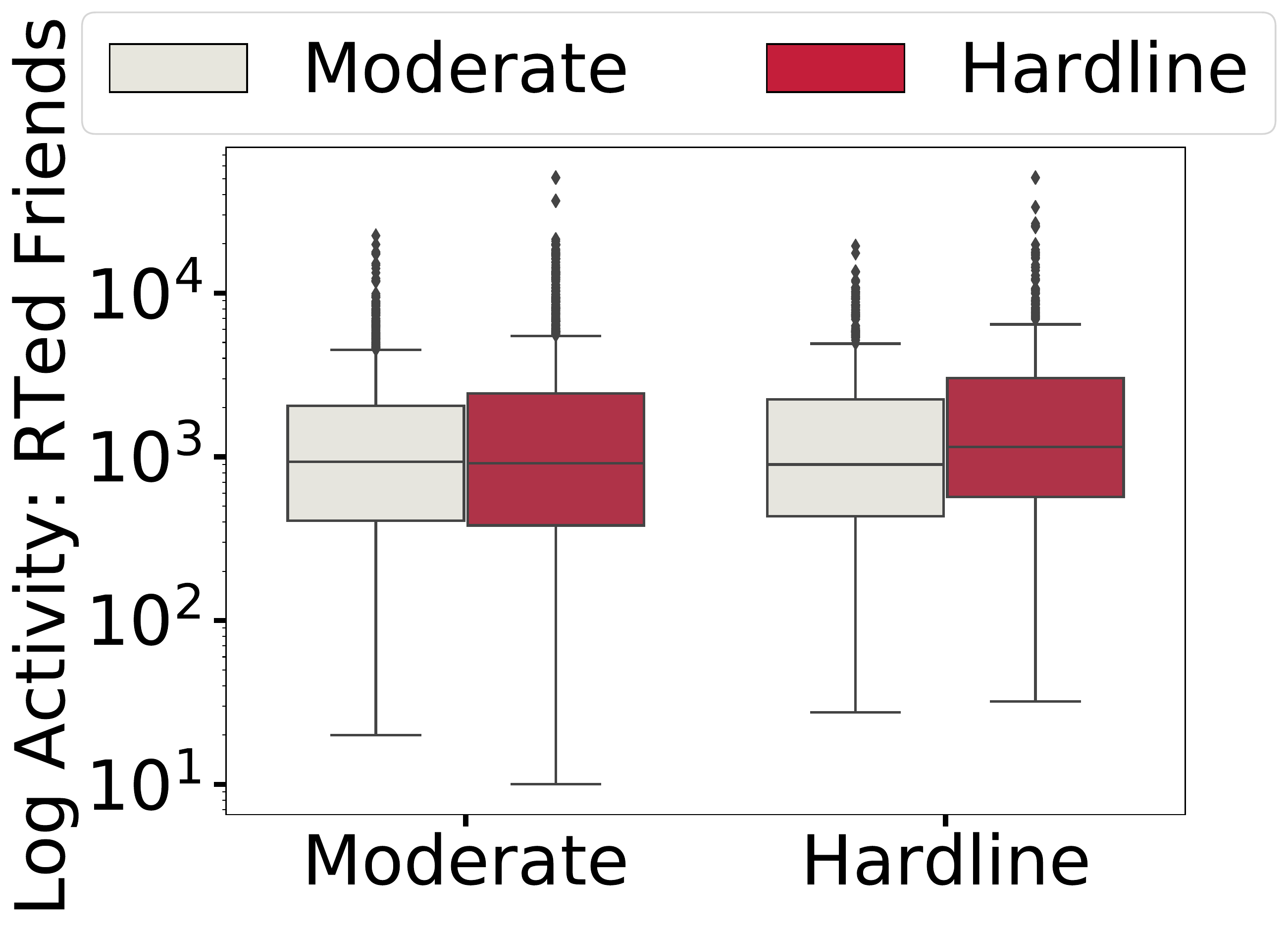}}
 \caption{Comparing Activity. Despite retweeted follower-graph friends having a higher activity than the ones who aren't retweeted, hardline retweeted friends of both moderate and hardline users tend to be most active.}
\label{fig:activity}
\end{figure}

\begin{figure}[th]
    \centering
    \includegraphics[width=0.98\linewidth]{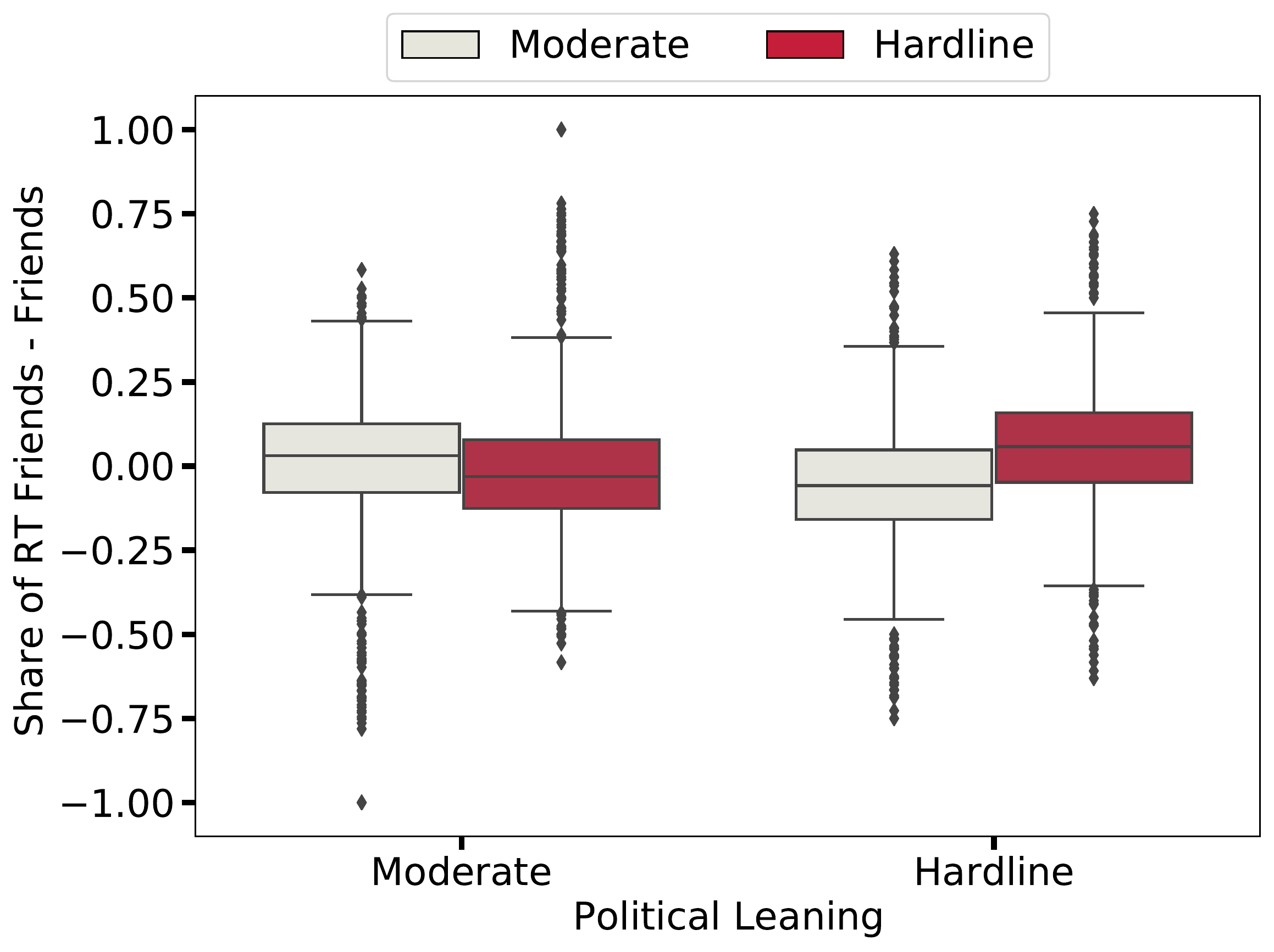}
    \caption{Homophily explains echo chambers. Individuals prefer to surround themselves with others who share similar polarities. Most moderates and hardliners have a higher fraction of their retweeted follower-graph friends to be of the same ideology as them in comparison to the follower-graph friends that they don't retweet.}
    \label{fig:fraction_by_group}
\end{figure}

\subsection{Origins of Bias}
Why are retweet friends more hardline than follower-graph friends?
During the time of data collection, Twitter was not yet personalizing timelines, so algorithmic amplification of ideologically congruent content~\cite{chen2021neutral} does not explain the observed effect. 
We therefore look at alternate explanations. Are follower-graph friends who are retweeted more active than follower-graph friends who are not retweeted? If so, are the retweeted follower-graph friends who are most similar to the user in moderacy more active than dissimilar friends? If the answers to these questions are yes, then it is likely that their content is more salient and therefore gets more attention. 

\subsubsection{Are Polarized Users More Active?}
A potential explanation as to why hardline seed users retweet more hardline friends and moderate users retweet more moderate friends 
(Fig. \ref{fig:exposure_vs_sharing}) is that follower-graph friends who are retweeted are more active than friends who are not retweeted. 

We find that the 
bias cannot be explained by friend activity alone. To demonstrate, we first compare the activity of follower-graph friends who are retweeted and follower-graph friends who aren't retweeted. We find that retweeted friends are more active than ones who aren't retweeted, a result that is statistically significant  under the Mann-Whitney U Test ($p<0.001$) (Refer Fig. \ref{fig:activity}(a)). 
However, there is no difference in the activity of retweeted friends based on their moderacy scores Fig.~\ref{fig:activity}(b).

\subsubsection{Selective Attention to Ideologically Polarized Content}

To understand why retweet exposures amplify the echo chamber effect, it is important to understand who is paying attention to whom. To measure this, for each moderacy group, we calculate the fraction of moderates and hardliners amongst users' retweeted follower-graph friends and not-retweeted follower-graph friends. For each user, we then compute the difference of these two fractions. 

Fig. \ref{fig:fraction_by_group} shows the boxplot of differences for moderate and hardline users. For moderate users, we see that a higher fraction of their retweeted follower-graph friends are also moderate in comparison to the ones they did not retweet. This effect is more pronounced for hardline users. The differences shown in Fig. \ref{fig:fraction_by_group} are statistically significant ($p<0.001$). On average, $69\%$ of moderate users' retweeted friends are also moderate as opposed to $66\%$ amongst the friends they don't retweet. For hardline users, $49\%$ of their retweeted friends are also hardline as opposed to $31\%$ amongst the ones they do not retweet.

\section{Conclusions}
Network connections expose people to information in online social media. When people link to similar others in social networks, they risk embedding themselves within echo chambers that expose them to similar views and insulate them from opposing viewpoints. Exposures have been studied by analyzing the accounts users follow. An alternative approach  to analyze retweet interactions thereby capturing what users pay attention to. However, it is not generally known how well these representations of exposure agree. Leveraging political polarities at the scale of pay-level domains, we quantify users' moderacy, as well as the moderacy of their friends. We then compare our estimates of information exposure via the follower and retweet networks. Relying on the retweet network to measure the information people see systematically amplifies its polarization compared to what users see their friends in the follower network post. This reveals several key insights into the nature of information exposure in online environments.

We find a significant correlation between a user's polarization and the polarization of the information they see in both networks, which points to the existence of echo chambers, i.e., ideologically similar friends who expose users to information that aligns with users' own attitudes. However, we find that retweet graph consistently amplifies true exposures. Retweet exposures are more correlated with individual moderacy scores in comparison to follower-graph exposures. Follower-graph exposures are also less varied in comparison retweet-graph exposures with the latter having similar variance to individual moderacy scores.  We find that under retweet exposures, both hardline and moderate users see more hardline and moderate content respectively, than they would under follower-graph exposures. The existence of an amplified echo chamber effect with retweet exposures is also made evident by comparing the entropy of follower-graph and retweet-graph friends. The differences between what users follow and retweet highlight that looking at retweet relationships to quantify exposures may in fact amplify the echo chamber effect.    
 
We find that as users become more hardline or moderate, their retweet-graph friends expose them to more hardline or moderate content, respectively, than they would see from their follower-graph friends. Therefore, exposures computed through the follower graph underestimate the extent to which individuals pay attention to ideologically congruent information. We also find that users pay more attention to friends who are ideologically similar to them. 
As we increase how much attention users pay to friends (by varying the retweet threshold), we find that exposure bias increases.

Finally, we investigate the reason behind retweet exposures being more extreme. Given that the dataset was collected before November 2014 the distortions cannot be attributed to algorithmic curation of timelines. We then assess if retweeted follower-graph friends are more active than ones who weren't retweeted and if this translates into an increased availability of ideologically congruent content for users to retweet. While, retweeted-friends were more active,  hardline ones amongst them had the highest activity. This does not explain why moderate users see more moderate content through retweet exposures. 

By focusing on not just \textit{who was retweeted} but factoring in \textit{who made the retweet} we find that, individuals prefer to retweet others who are at least as polarized as they are. We see that a large proportion of retweets are of others who are at least as polarized as the user. We therefore argue that by relying on retweet relationships to understand the echo chamber effect, previous studies may have relied on a biased estimate of exposures. These results point to important considerations for researchers studying polarization through the lens of social media. Studies should factor in user attention span when quantifying exposures. Owing to the proliferation of content it is impractical to assume that users can pay any attention if not equal attention to all their friends. Repeated interactions between users can be a viable proxy to assess what users pay attention to. The fact that these preferences existed prior to the introduction of personalization algorithms highlights an inherence in user behavior which can be exacerbated by content curation. 

\subsection{Limitations and Future Work}
Our study has the following limitations which should be taken into account. The set of seed users whose friendship links we collected have a strong liberal bias. We mitigate the bias by focusing on the moderacy axis which represents the intensity of ideological belief and not ideology itself. Additionally, we only look at tweets and retweets made by seed users and their friends. While we have data about the friends of seed users we don't have data about the friends of friends. This creates sinks in the network because two friends of a seed user can have a link between them if one of them is also a seed user. In addition, any study of exposure done today would have to account for Twitter personalization algorithms, which may further exacerbate exposures. Media Bias-Fact Check by no means provides an all encompassing list of Pay-Level Domains and one can explore other sources such as NewsGuard, Adfontes Media etc. A comparison of these exposures with exposures from personalized timelines today is an interesting and important direction for future research. 

\subsection{Ethical Considerations}
The study was reviewed by our institutional review board and determined to be exempt. The Twitter data that we collect is only representative of English language speakers in the United States and is by no means representative of the general population. One way we can overcome this is by building multi-lingual approaches and probably expanding to tweets from across the world. Another ethical shortcoming could we the implication that findings on Twitter reflect real world phenomena. Twitter tends to be biased towards liberals and is mostly used by the younger population thereby not being representative of the real world. Political sharing behaviors are personal to the individual and in order to preserve anonymity we remove screen names from tweets.

\bibliographystyle{IEEEtran}
\bibliography{references}

\end{document}